\documentclass[12pt,journal]{IEEEtran}
\usepackage{epsfig}
\usepackage{amssymb}
\usepackage{graphicx}
\usepackage{graphics}
\usepackage{amsmath}
\usepackage{setspace}
\onecolumn
\newtheorem{theorem}{Theorem}
      \newtheorem{lemma}{Lemma}
      
\title{
Joint Source-Channel Coding on a Multiple Access Channel with Side Information}

\author{R Rajesh, \thanks{Preliminary versions of parts of this paper appear  in WCNC 06, ISIT 08 and ISITA 08. 
R Rajesh  and Vinod Sharma are with the Dept of Electrical Communication Engineering, Indian Institute of Science, Bangalore, India. V K Varshneya is with IBM India Research Lab, Bangalore, India. Email: rajesh@pal.ece.iisc.ernet.in, vinod@ece.iisc.ernet.in, virendra@pal.ece.iisc.ernet.in. This work is partially supported by DRDO-IISc program on advanced research in mathematical engineering.}
\and  Vinod Sharma \IEEEmembership{Senior Member~IEEE} and \and V K Varshneya}

\begin{document}
\maketitle
\thispagestyle{empty}
\pagestyle{empty}
\begin{abstract}
We consider the problem of transmission of several distributed correlated sources over a multiple access channel (MAC) with side information at the sources and the decoder. Source-channel separation does not hold for this channel. Sufficient conditions are provided for transmission of sources with a given distortion. The source and/or the channel could have continuous alphabets (thus Gaussian sources and Gaussian MACs are special cases). Various previous results are obtained as special cases. We also provide several good joint source-channel coding schemes for discrete sources and discrete/continuous alphabet channel. 

\end{abstract}
\section{Introduction and Survey}\label{sec1.1}
In this paper we consider the transmission of information from several correlated sources over a multiple access channel  with side information. This system does not satisfy source-channel separation (\cite{Cover80multiple}). Thus for optimal transmission one needs to consider joint source-channel coding. We will provide several good joint source-channel coding schemes. 

Although this topic has been studied for last several decades, one recent motivation is the problem of estimating a random field via sensor networks. Sensor nodes have limited computational and storage capabilities and very limited energy \cite{akylidiz02survey}. These sensor nodes need to transmit their observations to a fusion center which uses this data to estimate the sensed random field. Since transmission is very energy intensive, it is important to minimize it.

The proximity of the sensing nodes to each other induces high correlations between the observations of adjacent sensors. One can exploit these correlations to compress the transmitted data significantly (\cite{akylidiz02survey},~\cite{Baek04minimizing}). Furthermore, some of the nodes can be more powerful and  act as cluster heads (\cite{Baek04minimizing}). Nodes  transmit their data to a nearby cluster head which can further compress information before transmission to the fusion center. Transmission of data from sensor nodes to their cluster-head requires sharing the wireless multiple access channel (MAC). At the fusion center the underlying physical process is estimated. The main trade-off possible is between the rates at which the sensors send their observations and the distortion incurred in the estimation at the fusion center. The availability of side information at the encoders and/or the decoder can reduce the rate of transmission (\cite{gasp},~\cite{wyner}).

The above considerations open up new interesting problems in multi-user information theory and the quest for finding the optimal performance for various models of sources, channels and side information have made this an active area of research. The optimal solution is not known except in a few simple cases. In this paper a joint source channel coding approach is discussed under various assumptions on side information and distortion criteria. Sufficient conditions for transmission of discrete/continuous alphabet sources with a given distortion over a discrete/continuous  alphabet MAC are provided. These results generalize the previous results available on this problem.

In the following we survey the related literature. Ahlswede \cite{ash} and Liao \cite{liao} obtained the capacity region of a discrete memoryless MAC with independent inputs. Cover, El Gamal and Salehi \cite{Cover80multiple} made further significant progress by providing sufficient conditions for transmitting losslessly correlated observations over a MAC. They proposed a `correlation preserving' scheme for transmitting the sources. This mapping is extended to a more general system with several principle sources and several side information sources subject to cross observations at the encoders in \cite{han}. However single letter characterization of the capacity region is still unknown. Indeed Duek \cite{Duek81note} proved that the conditions given in \cite{Cover80multiple} are only sufficient and may not be necessary. In  \cite{kang} a finite letter upper bound for the problem is obtained. It is also shown in \cite{Cover80multiple} that the source-channel separation does not hold in this case.  The authors of  \cite{Medrad06seperation} obtain a condition for separation to hold in a multiple access channel. 

The capacity region for the distributed lossless source coding problem for correlated sources is given in the classic paper by Slepian and Wolf (\cite{slepian}). Cover (\cite{cov}) extended Slepian-Wolf results to an arbitrary number of discrete, ergodic sources using a technique called `random binning'. Other related papers on this problem are \cite{han},~\cite{serv}.

Inspired by Slepian-Wolf results, Wyner and Ziv \cite{wyner} obtained the rate distortion function for source coding with side information at the decoder. It is shown that the knowledge of  side information at the encoders in addition to the decoder, permits the transmission  at a lower  rate. This is in contrast to the lossless case considered by Slepian and Wolf. The rate distortion function when encoder and decoder both have side information was first obtained by Gray (See \cite{tb}). Related work on side information coding is \cite{wornell},~\cite{draper},~\cite{pr}. The lossy version of Slepian-Wolf problem is called multi-terminal source coding problem and despite numerous attempts (e.g., ~\cite{by},~\cite{Oohama97Gaussian}) the exact rate region is not known except for a few special cases. First major advancement was in  Berger and Tung (\cite{tb}) where an inner and an outer bound on the rate distortion region was obtained. Lossy coding of continuous sources at the  high resolution limit is studied in \cite{zamir} where an explicit single-letter bound is obtained. Gastpar (\cite{gasp}) derived an inner and an outer bound  with decoder side information and proved the tightness of his bounds  when the sources are conditionally independent given the side information. The authors in \cite{ncc} obtain inner and outer bounds on the rate region with side information at the encoders and the decoder. In \cite{ong} an achievable rate region for a MAC with correlated sources and feedback is given.

 
The distributed Gaussian source coding problem is discussed in \cite{Oohama97Gaussian},~\cite{Wagner05rate}. For two users exact rate region is provided in  \cite{Wagner05rate}. The capacity of a Gaussian MAC (GMAC) for independent sources with feedback is given in \cite{Ozarow84capacity}. In \cite{Lapidoth01sending} one necessary and two sufficient conditions for transmitting a bivariate jointly Gaussian source over a GMAC are provided. It is shown that the amplify and forward scheme is optimal below  a certain SNR. The performance comparison of the schemes given in \cite{Lapidoth01sending} with a separation-based scheme is given in \cite{Rajesh07allerton}. GMAC under received power constraints is studied in \cite{Gastpar04gaussian} and it is shown that the source-channel separation holds in this case. 

In~\cite{Gastpar03source} the authors discuss a joint source channel coding scheme over a MAC and show the scaling behavior for the Gaussian channel. A Gaussian sensor network in distributed and collaborative setting is studied in \cite{Iswar05rate}. The authors show that it is better to compress the local estimates than to compress the raw data. The scaling laws for a many-to-one data-gathering channel are discussed in \cite{elgmal}. It is shown that the transport capacity of the network scales as $\mathcal{O}(logN)$ when the number of sensors $N$ grows to infinity and the total average power remains fixed. The scaling laws for the problem without side information are also discussed in \cite{Gastpar05power} and it is shown that  separating source coding from channel coding may require exponential growth, as a function of number of sensors, in communication bandwidth. A lower bound on best achievable distortion as a function of the number of sensors, total transmit power, the  degrees of freedom of the underlying process and the spatio-temporal communication bandwidth is given.

 The joint source-channel coding problem also bears relationship to the CEO problem \cite{ceo}. In this problem, multiple encoders observe different, noisy versions of a single information source and communicate it to a single decoder called the CEO which is required to reconstruct the source within a certain distortion. The Gaussian version of the CEO problem is studied in \cite{Oohama98rate}. 

This paper makes the following contributions. It obtains sufficient conditions for transmission  of correlated sources  with given distortions over a MAC with side information. The source/channel alphabets can be discrete or continuous. The sufficient conditions are strong enough that previous known results are special cases. Next we  obtain a bit to Gaussian mapping which provides correlated Gaussian channel codewords for discrete distributed sources.

The paper is organized as follows. Sufficient conditions for transmission of distributed sources over a MAC with side information and given distortion are obtained in Section~\ref{sec1.3}. The sources and the channel alphabets can be continuous or discrete. Several previous results are recovered as special cases in Section~\ref{sec1.4}. Section~\ref{sec1.5} considers the important case of transmission of discrete correlated sources over a GMAC and presents  a  new joint source-channel coding scheme. Section~\ref{sec1.6} briefly considers Gaussian sources over a GMAC. Section~\ref{sec1.12} concludes the paper. The proof of the main theorem is given in Appendix A. The proofs of several other results are provided in later appendices.

\section{Transmission of correlated sources over a MAC}\label{sec1.3}
\begin{figure}[h]
\centering
\includegraphics [width=7.5cm,height=4.2 cm] {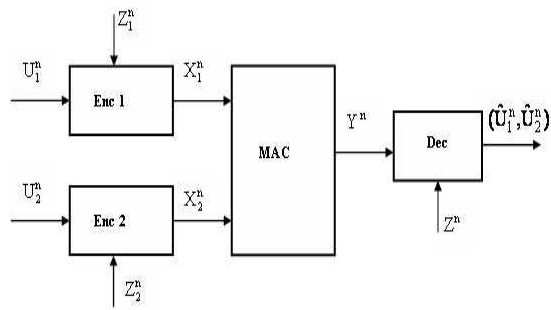}
\caption{Transmission of correlated sources over a MAC with side information.}
\label{fig1.1}
\end{figure}

We consider the transmission of memoryless dependent sources, through a memoryless multiple access channel  (Fig.~\ref{fig1.1}). The sources and/or the channel input/output alphabets can be discrete or continuous. Furthermore, side information about the transmitted information may be available at the encoders and the decoder. Thus our system is very general and covers many systems studied earlier.

Initially we consider two sources $(U_1,U_2)$ and side information random variables $Z_1, Z_2, Z$ with a known joint distribution $F(u_1, u_2, z_1, z_2, z)$. Side information $Z_i$ is available to encoder $i,~i= {1, 2}$ and the decoder has side information $Z$. The random vector sequence $\{(U_{1n}, U_{2n}, Z_{1n}, Z_{2n}, Z_n), n\geq1\}$ formed from the source outputs and the side information with distribution $F$ is independent identically distributed ({\it{iid}}) in time. We will denote $\{U_{1k},~k=1,...,n\}$ by $U_1^n$. Similarly for other sequences.  The sources transmit their codewords $X_{in}$'s to a single decoder through a memoryless multiple access channel. The channel output $Y$ has distribution $p(y|x_1,x_2)$ if $x_1$ and $x_2$ are transmitted at that time. Thus, $\{Y_n\}$ and $\{X_{1n}, X_{2n}\}$ satisfy $p(y_k|y^{k-1},x_1^k,x_2^k)=p(y_k|x_{1k},x_{2k})$. The decoder receives $Y_n$ and also has access to the side information $Z_n$. The encoders at the two users do not communicate with each other except via the side information. The decoder uses the channel outputs and its side information  to estimate the sensor observations $U_{in}$   as $ \hat{U}_{in},~i= {1, 2}$. It is of interest to find encoders and a decoder such that  $\{U_{1n},U_{2n},n\geq1\}$ can be transmitted over the given MAC with $E[d_1(U_1,\hat{U_1})]\leq D_1$ and  $E[d_2(U_2,\hat{U_2})]\leq D_2$
 where $d_i$  are non-negative distortion measures and $D_i$ are the given distortion constraints. If the distortion measures are unbounded we assume that there exist $u_i^*$ such that $E[d_i(U_i,u_i^*)] < \infty,~i= {1, 2}$. This covers the important special case of mean square error (MSE) if $E[U_i^2] < \infty,~i=1,2$.
 
 Source channel separation does not hold in this case.

For discrete sources a common distortion measure is Hamming distance,
\begin{eqnarray*}
d(x,x')= \begin{cases}
1,~& \text{if $x \ne x'$},\\
0,~& \text{if $x = x'$}.
\end{cases}
\end{eqnarray*}
For continuous alphabet sources the most common distortion measure is $d(x,x')=(x-x')^2$. To obtain the results for lossless case from our Theorem 1 below, we assume that $d_i(x,x')=0 \Leftrightarrow x=x'$, e.g., Hamming distance.

$Definition$: The source $(U_1^n ,U_2^n)$ can be transmitted over the multiple access channel with distortions ${\bf{D}}{\buildrel\Delta \over=}(D_1,D_2)$ if for any $\epsilon  > 0 $ there is an $n_0$ such that for all $n > n_0$  there exist encoders $f_{E,i}^n: \mathcal{U}_i^n \times \mathcal{Z}_i^n \rightarrow \mathcal{X}_i^n , i ={1,2}$ and a decoder ${{f}_{D}^n}: \mathcal{Y}^n \times \mathcal{Z}^n \rightarrow (\mathcal{\hat{U}}_1^n , \mathcal{\hat{U}}_2^n) $   such that $ \frac{1}{n}E\left[\sum_{j=1}^nd (U_{ij},\hat{U}_{ij})\right]\leq D_i+ \epsilon,~i={1,2}$ where $(\hat{U}_1^n,\hat{U}_2^n )= f_D(Y^n,Z^n)$ and   $\mathcal{U}_i,~\mathcal{Z}_i,~\mathcal{Z},~\mathcal{X}_i,~\mathcal{Y},~\hat{\mathcal{U}_i} $  are the sets in which $ U_i,~Z_i,~Z,~X_i,~Y,~\hat{U}_i$ take values.

We denote the joint distribution of $(U_1, U_2)$ by $p(u_1,u_2)$. Also, $X\leftrightarrow Y \leftrightarrow Z$   will denote that $\{X, Y, Z\}$ form a Markov chain.

Now we state the main Theorem. 
\begin{theorem}
\label{theorem1}
    	 A source  can be transmitted over the multiple access channel with distortions $(D_1, D_2)$ if there exist random variables $(W_1, W_2,  X_1,  X_2)$ such  that
\begin{eqnarray*}
(1)~ p(u_1,u_2,z_1,z_2,z,w_1,w_2,x_1,x_2,y) = 
      p(u_1,u_2,z_1,z_2,z)
 p(w_1|u_1,z_1)  p(w_2|u_2,z_2).~~~~~~~~~~~~~~~~~~~~~~~~~~~~~~~~~~~~~~~~~~~~~~ \\  
      p(x_1|w_1)p(x_2|w_2)p(y|x_1,x_2)~~~~~~~~~~~~~~~~~~~~~~~~~~~~~~~~~~~~~~~~~~~~~~  
\end{eqnarray*}
and\\ 
(2) there exists a function $f_D: \mathcal{W}_1 \times \mathcal{W}_2 \times \mathcal{Z} \rightarrow (\hat{\mathcal{U}}_1 \times  \hat{\mathcal{U}}_2) $  such that $ E[d(U_{i},\hat{U}_{i})]\leq D_i,~i=1,2$, where $(\hat{U}_1,\hat{U}_2)=f_D(W_1,W_2,Z)$ and the constraints
\begin{eqnarray}
I (U_1,Z_1; W_1 | W_2,Z)  &<&  I (X_1; Y | X_2, W_2, Z),\nonumber\\
I (U_2,Z_2; W_2 | W_1,Z) &<& I (X_2; Y | X_1, W_1, Z),\label{eqn1.1}\\	   		    
I (U_1,U_2,Z_1,Z_2 ; W_1, W_2 |Z) &<& I (X_1, X_2; Y |Z)\nonumber
\end{eqnarray}
are satisfied where $\mathcal{W}_i$  are the sets in which $ W_i$ take values.
\end{theorem}
$Proof$: See Appendix A. ~~~~~~~~~~~~~~~~~~~~~~~~~~~~~~~~~~~~~~~~~~~~~~~~~~~~~~~~~~~~~~~~~~~~~~~~~~~~~~~~~~~~~~~~~~~~~~~$\blacksquare$

In the proof of Theorem~\ref{theorem1} the encoding scheme involves distributed vector quantization $(W_1^n,W_2^n)$ of the sources $(U_1^n,U_2^n)$ and the side information $Z_1^n,Z_2^n$ followed by a correlation preserving mapping to the channel codewords $(X_1^n,X_2^n)$. The decoding approach involves first decoding $(W_1^n,W_2^n)$ and then obtaining the estimates $(\hat{U}_1^n,\hat{U}_2^n)$ as a function of $(W_1^n,W_2^n)$ and the decoder side information $Z^n$.

If the channel alphabets are continuous (e.g., GMAC) then in addition to the conditions in Theorem~\ref{theorem1} certain power constraints $E[X_i^2] \le P_i,~i=1,2$ are also needed. In general, we could impose a more general constraint $E[g_i(X_i)] \le \alpha_i$ where $g_i$ is some non-negative cost function. Furthermore, for continuous alphabet r.v.s (sources/channel input/output) we will assume that probability density exists so that one can use differential entropy (more general cases can be handled but for simplicity we will ignore them).

The dependence in $(U_1,U_2)$ is used in two ways in \eqref{eqn1.1}: to $reduce$ the quantities on the left and to increase the quantities on the right. The side information $Z_1$ and $Z_2$ effectively $increases$ the dependence in the inputs. 

If the source-channel separation holds then one can consider the capacity region of the channel. For example, when there is no side information $Z_1,Z_2,Z$ and the sources are independent then we obtain the rate region
\begin{eqnarray}
R_1 \leq I(X_1;Y|X_2),~
R_2 \leq I(X_2;Y|X_1),~\label{ne}
R_1+ R_2 \leq I(X_1,X_2;Y).
\end{eqnarray}
This is the well known rate region of a  MAC (\cite{Cover04elements}). To obtain \eqref{ne} from \eqref{eqn1.1}, take $(Z_1,Z_2,Z)$ independent of $(U_1,U_2)$. Also, take $U_1,U_2$ discrete, $W_i=U_i$ and $X_i$ independent of $U_i,~i=1,~2$. 

In Theorem~\ref{theorem1} it is possible to include other distortion constraints. For example, in addition to the bounds on $E[d(U_i,\hat{U}_i)]$ one may want a bound on the joint distortion $E[d((U_1,U_2),(\hat{U}_1,\hat{U}_2))]$. Then the only modification needed in the statement of the above theorem is to include this also as a condition in defining $f_D$.

If we only want to estimate a function $g(U_1,U_2)$ at the decoder and not $(U_1,U_2)$ themselves, then again one can use the techniques in proof of Theorem~\ref{theorem1} to obtain sufficient conditions. Depending upon $g$, the conditions needed may be weaker than those needed in \eqref{eqn1.1}. We will explore this in more detail in a later work.

In our problem setup the side information $Z_i$ can be included with source $U_i$ and then we can consider this problem as one with no side information at the encoders. However, the above formulation has the advantage that our conditions \eqref{eqn1.1} are explicit in $Z_i$.

The main problem in using Theorem~\ref{theorem1} is in obtaining good source-channel coding schemes providing $(W_1,W_2,X_1,X_2)$ which satisfy the conditions in the theorem for a given source $(U_1,U_2)$ and a channel. A substantial part of this paper will be devoted to this problem.

\subsection{ Extension to multiple sources} 

The above results can be generalized to the multiple $(\geq2)$ source   case. Let $\mathcal{S}={1,2,...,M}$ be the set of sources with joint distribution $p(u_1,...,u_M)$.
	
\begin{theorem}
Sources $(U_i^n ,i\in \mathcal{S})$  can be communicated in a distributed fashion over the memoryless multiple access channel $p(y|x_i ,i\in \mathcal{S})$  with distortions $(D_i ,i\in \mathcal{S})$  if there exist auxiliary random variables $(W_i,X_i ,i\in \mathcal{S})$ satisfying
\begin{eqnarray*}
(1)~ p(u_i,z_i,z,w_i,x_i,y ,i \in \mathcal{S}) =   
      p(u_i,z_i,z,i \in \mathcal{S})p(y|x_i,i \in \mathcal{S})
 \prod_{j \in \mathcal{S}}p(w_j|u_j,z_j)p(x_j|w_j),~~~~~~~~~~~~~~~~~ 
\end{eqnarray*}
(2) there exists a function $f_D: \prod_{j \in \mathcal{S}}\mathcal{W}_j \times  \mathcal{Z} \rightarrow (\hat{\mathcal{U}}_i,i \in \mathcal{S}) $  such that $ E[d(U_{i},\hat{U}_{i})]\leq D_i,~i \in \mathcal{S}$ and the constraints
\begin{equation}
I (U_A,Z_A; W_A | W_{A^c},Z) < I (X_A; Y | X_{A^c}, W_{A^c}, Z), ~\text{for all}~ A \subset \mathcal{S}
\label{eqn1.3}
\end{equation}
are satisfied where $U_A = (U_i,~i \in A)$, $A^c$ is the complement of set $A$ and similarly for other r.v.s (in case of continuous channel alphabets we also need the power constraints $E[X_i^2] \leq P_i,~i=1,...,|\mathcal{S}|)$.
\end{theorem}

\subsection{Example}
We provide an example  to show the reduction possible in transmission rates by exploiting the correlation between the sources, the side information and the permissible distortions.

Consider $(U_1, U_2)$ with the joint distribution: 
$P(U_1=0;U_2=0)= P(U_1=1;U_2=1)=1/3;  P(U_1=1;U_2=0)= P(U_1=0;U_2=1)=1/6.$
If we use independent encoders which do not exploit the correlation among the sources then we need $R_1 \geq H(U_1) = 1~bit$ and $R_2 \geq H (U_2) = 1~bit$ for lossless coding of the sources. If we use Slepian-Wolf coding (\cite{slepian}), then $R_1 \geq H(U_1|U_2) = 0.918~bits, R_2 \geq H (U_2|U_1) = 0.918~bits$ and $R_1+ R_2 \geq H(U_1, U_2) = 1.918~ bits$ suffice. 

Next consider a multiple access channel such that $Y = X_1+ X_2$ where $X_1$ and $X_2$ take values from the alphabet $\{0,1\}$ and $Y$  takes values from the alphabet $\{0,1,2\}$. This does not satisfy the separation conditions in \cite{Medrad06seperation}. The sum capacity $C$ of such a channel with independent $X_1$ and $X_2$ is $1.5~ bits$ and if we use source-channel separation, the given sources cannot be transmitted losslessly because $ H(U_1, U_2) > C$. Now we use a joint source-channel code to improve the capacity of the channel. Take $X_1=U_1$ and $X_2=U_2$. Then the sum rate capacity of the channel is improved to $I(X_1,X_2;Y)=1.585~bits$. This is still not enough to transmit the sources over the given MAC. Next we exploit the side information.

Let the side-information random variables be generated as follows. $Z_1$ is transmitted from source 2 by using a (low rate) binary symmetric channel (BSC) with cross over probability $p=0.3$. Similarly $Z_2$ is transmitted  from source 1 via a similar BSC. Let $Z = (Z_1,Z_2,V)$, where $V= U_1 .U_2 .N$, $N$ is a binary random variable with $P(N=0) = P(N=1) = 0.5 $ independent of $U_1$ and $U_2$ and  `.' denotes the logical AND operation. This  denotes the case when the decoder has access to the encoder side information and also has some extra side information. Then from \eqref{eqn1.1} if we use just the side information $Z_1$ the sum rate for the sources needs to be $1.8~ bits$. By symmetry the same holds if we only have $Z_2$. If we use $Z_1$ and $Z_2$ then we can use the sum rate $1.683~ bits$. If only $V$ is used then the sum rate needed is $1.606~ bits$. So far we can still not transmit $(U_1,U_2)$ losslessly if we use the coding $U_i=X_i,~i=1,2$. If all the information in $Z_1, Z_2,V$ is used then we need $R_1+ R_2 \geq 1.4120 ~bits$. Thus with the aid of $Z_1, Z_2, Z$ we can transmit $(U_1, U_2)$ losslessly over the MAC even with independent $X_1$ and $X_2$.

Next we consider the distortion criterion to be the Hamming distance and the allowable distortion as 4\%. Then for compressing the individual sources without side information we need  $R_i \geq H(p)-H(d) = 0.758~ bits,~i=1,2$, where $H(x)=-xlog_2(x)-(1-x)log_2(1-x)$. Thus we still cannot transmit $(U_1, U_2)$ with this distortion when $(X_1, X_2)$ are independent. Next assume the side information $Z=(Z_1,Z_2)$ to be available at the decoder only. Then we need $R_1 \geq  I(U_1;W_1)- I(Z_1; W_1)$ where $W_1$ is an auxiliary random variable generated from $U_1$.  This implies that $R_1 \geq 0.6577~ bits$ and $R_2 \geq 0.6577~ bits$ and we can transmit with independent $X_1$ and $X_2$.
\section{Special Cases}\label{sec1.4}

In the following we show that our result contains several  previous studies as special cases. The practically important special case of GMAC will be studied in detail in later sections. There we will discuss several specific joint source-channel coding schemes for GMAC and compare their performance.

\subsection{Lossless multiple access communication with correlated sources}
Take   $(Z_1,Z_2,Z) \bot (U_1, U_2)$ ($X \bot Y$ denotes that r.v. $X$ is independent of r.v. $Y$)  and $W_1=U_1$ and $W_2=U_2$ where $U_1,U_2$ are discrete sources. Then the constraints of \eqref{eqn1.1} reduce to 
\begin{gather}
H (U_1| U_2)  < I (X_1; Y | X_2, U_2),~
H(U_2 | U_1)  < I (X_2; Y | X_1, U_1),~\label{covereq}	   		    
H(U_1,U_2)  <  I (X_1, X_2; Y)
\end{gather}
where  $X_1\leftrightarrow U_1 \leftrightarrow U_2\leftrightarrow X_2$. These are the conditions obtained in  \cite{Cover80multiple}. 

If $U_1,~U_2$ are independent, then $H(U_1|U_2)=H(U_1)$ and $I(X_1;Y|X_2,U_2)=I(X_1;Y|X_2)$.
\subsection{Lossy multiple access communication}
Take  $(Z_1,Z_2,Z) \bot (U_1, U_2, W_1, W_2)$ . In this case the constraints in \eqref{eqn1.1} reduce to
\begin{gather}
   I (U_1 ;W_1 | W_2)   <  I (X_1;Y |X_2, W_2),~
   I (U_2; W_2| W_1)   <  I (X_2;Y |X_1, W_1),\nonumber\\			  	 
 I (U_1,U_2 ;W_1,W_2)   <  I (X_1,X_2;Y).
\label{eqna}
\end{gather}
This is an immediate generalization of \cite{Cover80multiple} to the lossy case.
\subsection{Lossless multiple access communication with common information}
Consider $U_1=(U_1',U_0'),~U_2=(U_2',U_0')$ where $U_0',U_1',U_2'$ are independent of each other. $U_0'$ is interpreted as the common information at the two encoders. Then, taking $(Z_1,Z_2,Z)\bot (U_1,U_2)$, $W_1=U_1$ and $W_2=U_2$ we obtain sufficient conditions for lossless transmission as
\begin{gather}
   H(U_1')   <  I (X_1;Y | X_2, U_0'),~
    H(U_2')   <  I (X_2;Y |X_1, U_0'),\nonumber\\			  	 
 H(U_1')+H(U_2')+H(U_0')   <  I (X_1,X_2;Y).
\label{eqnloss}
\end{gather}
This provides the capacity region of the MAC with common information available in \cite{slep}. 

Our results generalize this result to lossy transmission also.
\subsection{Lossy distributed source coding with side information}
The multiple access channel is taken as a dummy channel which reproduces its inputs. In this case we obtain that the sources can be coded with rates $R_1$ and $R_2$ to obtain the specified distortions at the decoder if  
\begin{gather}
R_1  >  I (U_1,Z_1; W_1 | W_2,  Z),~
R_2  >   I (U_2,Z_2; W_2 | W_1,  Z),\nonumber\\			     	
R_1  + R_2 >  I (U_1, U_2,Z_1,Z_2 ; W_1, W_2 | Z)
\label{eqnb}
\end{gather}
where $R_1,~R_2$ are obtained by taking $X_1 \bot X_2$.

This recovers the result in \cite{ncc}, and generalizes the results in ~\cite{gasp},~\cite{slepian},~\cite{wyner}.

\subsection{Correlated sources with lossless transmission over MAC with receiver side information}

If we consider $(Z_1,Z_2)\bot (U_1,U_2)$, $W_1=U_1$ and $W_2=U_2$ then we recover the conditions 
\begin{gather}
H(U_1|U_2,Z)  <  I (X_1; Y | X_2, U_2, Z),~
H(U_2|U_1,Z)  < I (X_2; Y | X_1, U_1, Z),\nonumber\\	   		    
H (U_1,U_2 |Z) < I (X_1, X_2; Y |Z)
\label{eqn1.4}
\end{gather}
in $Theorem~2.1$ in \cite{elza}.

\subsection{ Mixed Side Information }
The aim is to determine the rate distortion function for transmitting a source $X$ with the aid of side information $(Y,Z)$ (system in Fig 1(c) of \cite{effros}). The encoder is provided with $Y$ and the decoder has access to both $Y$ and $Z$. This represents the Mixed side information (MSI) system which combines the conditional rate distortion system and the Wyner-Ziv system. This has the system in Fig 1(a) and (b) of \cite{effros} as special cases. 

The results of Fig 1(c) can be recovered from our Theorem if we take $X,Y,Z,W$ in \cite{effros} as $U_1=X,Z=(Z,Y),Z_1=Y$ and $W_1=W$. We also take $U_2$ and $Z_2$  to be  constants. The acceptable rate region is given by $R > I(X ; W|Y,Z)$, where $W$ is a random variable with the property $W\leftrightarrow (X,Y) \leftrightarrow Z$ and for which there exists a decoder function such that the distortion constraints are met.

\subsection{ Compound MAC and Interference channel with side information}
In compound MAC sources $U_1$ and $U_2$ are transmitted through a MAC which  has two outputs $Y_1$ and $Y_2$. Decoder $i$ is provided with $Y_i$ and $Z_i,~i=1,2$. Each decoder is supposed to reconstruct both the sources. We take $W_1=U_1$ and $W_2=U_2$. We can consider this system as two MAC's. Applying  \eqref{eqn1.1} twice
we have for $i=1,2$,
\begin{gather}
H(U_1|U_2,Z_i)  <  I (X_1; Y_i | X_2, U_2, Z_i),~
H(U_2|U_1,Z_i)  < I (X_2; Y_i | X_1, U_1, Z_i),\nonumber\\	   		    
H (U_1,U_2 |Z_i) < I (X_1, X_2; Y_i |Z_i).
\label{eqn1mod}
\end{gather}

This recoves the achievability result in \cite{elza2}. This provides the achievability conditions in \cite{elza2} for strong  interference channel conditions also.

\subsection{ Correlated sources over orthogonal channels with side information}
The sources transmit their codewords $X_i$'s to a single decoder through memoryless orthogonal channels having transition probabilities $p(y_1|x_1)$ and $p(y_2|x_2)$. Hence in the theorem, $Y=(Y_1,Y_2)$ and $Y_1\leftrightarrow X_1 \leftrightarrow W_1\leftrightarrow (U_1,Z_1)\leftrightarrow (U_2,Z_2)\leftrightarrow W_2\leftrightarrow X_2\leftrightarrow Y_2$. In this case the constraints in \eqref{eqn1.1} reduce to
\begin{eqnarray}
\label{conints2}  
I (U_1,Z_1; W_1 | W_2,Z)  &<&  I (X_1; Y_1 | W_2,Z) \leq I(X_1;Y_1),\nonumber\\
I (U_2,Z_2;W_2 |W_1,Z)  &<& I (X_2; Y_2 | W_1,Z)\leq I(X_2;Y_2),\\ 	
I (U_1, U_2,Z_1,Z_2 ; W_1, W_2|Z)  &<&   I (X_1, X_2; Y_1,Y_2|Z)\leq I(X_1;Y_1)+I(X_2;Y_2)\nonumber. 
\end{eqnarray}

The outer bounds in \eqref{conints2} are attained if the channel codewords $(X_1,X_2)$ are independent of each other. Also, the distribution of $(X_1,X_2)$ maximizing these bounds are not dependent on the distribution of $(U_1,U_2)$.

Using Fano's inequality, for lossless transmission of discrete sources over discrete channels with side information, we can show that outer bounds in \eqref{conints2} are in fact necessary and sufficient. The proof of the converse is given in Appendix B.

If we take $W_1=U_1$ and $W_2=U_2$ and  the side information $(Z_1,Z_2,Z) \bot (U_1,U_2)$, we can recover the necessary and sufficient conditions in~\cite{barros}.

\subsection{ Gaussian sources over a Gaussian MAC}
Let $(U_{1}, U_{2})$ be jointly Gaussian with mean zero, variances $\sigma_i^2,~i=1,2$   and correlation $ \rho $. These sources have to be communicated over a Gaussian MAC with the output $Y_n$ at time $n$ given by $Y_n = X_{1n} + X_{2n} + N_n$  where $X_{1n}$  and $X_{2n}$  are the channel inputs at time $n$ and $N_n$  is a Gaussian random variable independent of $X_{1n}$  and  $X_{2n}$, with $E[N_n] =0 $ and $var(N_n)= \sigma_N^{2}$. The power constaints are $E[X_i^2] \le P_i,~i=1,2$. The distortion measure is the mean square error (MSE). We take $(Z_1,Z_2,Z) \bot (U_1,U_2)$.  We choose $W_1$ and $W_2$ according to the coding scheme given in \cite{Lapidoth01sending}. $X_1$ and $X_2$ are scaled versions of $W_1$ and $W_2$ respectively. Then from  \eqref{eqn1.1} we find that the rates at which $W_1$ and $W_2$ are encoded satisfy 

\begin{gather}
 R_1   \leq  0.5\log\left[ \frac{P_1}{{\sigma_N}^{2}}+ \frac{1}{(1- {\tilde{\rho}}^{2})}\right],~
    R_2   \leq  0.5\log\left[ \frac{P_2}{{\sigma_N}^{2}}+ \frac{1}{(1- {\tilde{\rho}}^{2})}\right],\nonumber \\
   R_1+R_2  \leq   0.5\log\left[ \frac{{\sigma_N}^{2}+ P_1 + P_2 + {2} {\tilde{\rho}}{\sqrt{P_1P_2}}} {{(1- {\tilde{\rho}}^{2})}{\sigma_N}^{2}} \right]. \label{const3}  
\end{gather}
where $\tilde{\rho}$ is the correlation between $X_1$ and $X_2$. The  distortions achieved are
\begin{eqnarray*}
D_1 &\geq&  var(U_1|W_1,W_2)= \frac{{\sigma_1}^{2}2^{-2R_1}\left[1-\rho^{2}\left(1-2^{-2R_2}\right)\right]}{(1- {\tilde{\rho}}^{2})},\\
D_2 &\geq&  var(U_2|W_1,W_2)= \frac{{\sigma_2}^{2}2^{-2R_2}\left[1-\rho^{2}\left(1-2^{-2R_1}\right)\right]}{(1- {\tilde{\rho}}^{2})}.
\end{eqnarray*}
This recovers the sufficient conditions  in \cite{Lapidoth01sending}.
\section{Discrete Alphabet Sources over Gaussian MAC}\label{sec1.5}
This system is practically very useful. For example, in a sensor network, the observations sensed by the sensor nodes are discretized and then transmitted over a GMAC. The physical proximity of the sensor nodes makes their observations correlated. This correlation can be exploited to compress the transmitted data and increase the channel capacity. We present a novel distributed `correlation preserving' joint source-channel coding scheme yielding jointly Gaussian channel codewords which transmit  the data efficiently over a GMAC.

Sufficient conditions for lossless transmission of two discrete correlated sources $(U_1,U_2)$ (generating $iid$ sequences in time) over a general MAC with no side information are obtained in \eqref{covereq}.

In this section, we further specialize these results to a  GMAC: $Y = X_1+ X_2+ N $ where $N$ is a Gaussian random variable independent of $X_1$ and $X_2$. The noise $N$ satisfies $E[N] = 0$ and $Var(N)=\sigma_N^2$ . We will also have the transmit power constraints: $E[X_i^2]\leq P_i, i=1,2$. Since source-channel separation does not hold for this system, a joint source-channel coding scheme is needed for optimal performance.

The dependence of right hand side (RHS) in \eqref{covereq} on input alphabets prevents us from getting a closed form expression for the admissibility criterion. Therefore we relax the conditions by taking away the dependence on the input alphabets to obtain good joint source-channel codes.
\begin{lemma}
 Under our assumptions, $ I(X_1; Y | X_2, U_2) \leq I (X_1; Y | X_2)$.
\label{lemma1}
\end{lemma}

$Proof$: See Appendix~\ref{l1}.~~~~~~~~~~~~~~~~~~~~~~~~~~~~~~~~~~~~~~~~~~~~~~~~~~~~~~~~~~~~~~~~~~~~~~~~~~~~~~~~~~~~~~~~~~~~~~~~~~~~~~~~~~~~~~~~$\blacksquare$ 
		            
Thus from \eqref{covereq},
\begin{eqnarray}
H (U_1 | U_2)  &<&  I (X_1;Y | X_2, U_2) \leq I ( X_1;Y | X_2 ),\label{sub1}\\
H (U_2 | U_1)  &<& I (X_2;Y | X_1, U_1) \leq  I ( X_2;Y | X_1),\label{sub2}\\
H (U_1, U_2)  &<&   I (X_1, X_2;Y)\label{sub3}.
\label{eqn0}
\end{eqnarray}
The relaxation of the upper bounds is only in \eqref{sub1} and \eqref{sub2} and not in \eqref{eqn0}.
 
We show that the relaxed upper bounds are maximized if $(X_1, X_2)$ is jointly Gaussian and the correlation $\rho$  between $X_1$ and $X_2$ is high (the highest possible $\rho$  may not give the largest upper bound  in \eqref{sub1}-\eqref{eqn0}). 

\begin{lemma}
A jointly Gaussian distribution for $(X_1, X_2)$ maximizes $I (X_1;Y | X_2)$, $I (X_2;Y | X_1)$ and $I (X_1, X_2;Y)$ simultaneously.
\label{lemma2}
\end{lemma}

$Proof$: See Appendix~\ref{l1}.~~~~~~~~~~~~~~~~~~~~~~~~~~~~~~~~~~~~~~~~~~~~~~~~~~~~~~~~~~~~~~~~~~~~~~~~~~~~~~~~~~~~~~~~~~~~~~~~~~~~~~~~~~~~~~~~$\blacksquare$ 

The difference between the bounds in  \eqref{sub1} is 
\begin{equation}
I(X_1,Y|X_2)-I(X_1,Y|X_2,U_2)=I(X_1+N;U_2|X_2).
\label{on}
\end{equation}
This difference is small if correlation between $(U_1,U_2)$ is small. In that case $H(U_1|U_2)$ and $H(U_2|U_1)$ will be large and \eqref{sub1}  and \eqref{sub2} can be active constraints. If correlation between $(U_1,U_2)$ is large, $H(U_1|U_2)$ and $H(U_2|U_1)$ will be small and \eqref{eqn0} will be the only active constraint. In this case the difference between the two bounds in \eqref{sub1} and \eqref{sub2} is large but not important. Thus, the outer bounds in \eqref{sub1} and \eqref{sub2} are close to the inner bounds whenever the constraints \eqref{sub1} and \eqref{sub2} are active. Often \eqref{eqn0} will be the only active constraint. 
 
Based on Lemma \ref{lemma2}, we use  jointly Gaussian channel inputs $(X_1, X_2)$ with the transmit power constraints. Thus we take $(X_1,X_2)$ with  mean vector $[0 ~~ 0]$ and covariance matrix $K_{X_1,X_2}=\\ \begin{pmatrix}
P_1 & \rho\sqrt{P_1P_2} \\
\rho\sqrt{P_1P_2} &  P_2
\end{pmatrix}$.
The outer bounds in \eqref{sub1}-\eqref{sub3} become 
$  0.5\log\left[ 1+ \frac{P_1(1- {{\rho}}^{2})} {{\sigma_N}^{2}}\right]$, 
 $ 0.5\log\left[ 1+ \frac{P_2(1- {{\rho}}^{2})} {{\sigma_N}^{2}}\right]$ and   			        
    $0.5\log$$\left[ 1+ \frac{P_1 + P_2 + {2} {{\rho}}{\sqrt{P_1P_2}}} {{\sigma_N}^{2}} \right]$ respectively.	   	
The first two upper bounds decrease as $\rho$  increases. But the third upper bound increases with $\rho$ and often the third constraint is the limiting constraint. Thus, once $(X_1,X_2)$ are obtained we can check for sufficient conditions \eqref{covereq}. If these are not satisfied for the $(X_1,X_2)$ obtained, we will increase the correlation $\rho$ between $(X_1,X_2)$ if possible (see details below). Increasing the correlation in $(X_1,X_2)$ will decrease the difference in \eqref{on} and increase the possibility of satisfying \eqref{covereq} when the outer bounds in \eqref{sub1} and \eqref{sub2} are satisfied. If not, we can increase ${\rho}$ further till we satisfy \eqref{covereq}.

The next lemma provides an upper bound on the correlation $\rho$  between $(X_1, X_2)$ possible in terms of the distribution of $(U_1,U_2)$.

\begin{lemma}
 Let $(U_1, U_2)$ be the correlated sources and $X_1\leftrightarrow U_1\leftrightarrow U_2\leftrightarrow X_2$ where $X_1$ and $X_2$ are jointly Gaussian. Then the correlation  ${\rho}$ between  $(X_1, X_2)$ satisfies $ {{\rho}}^2 \leq 1- 2^{-2I(U_1,U_2)}$.
 \label{lemma3}
 \end{lemma}
 
$Proof$: See Appendix~\ref{l1}.~~~~~~~~~~~~~~~~~~~~~~~~~~~~~~~~~~~~~~~~~~~~~~~~~~~~~~~~~~~~~~~~~~~~~~~~~~~~~~~~~~~~~~~~~~~~~~~~~~~~~~~~~~~~~~~~$\blacksquare$ 
 
It is stated in~\cite{Medrad06seperation}, without proof, that the correlation  between $(X_1, X_2)$ cannot be greater than the correlation of the source $(U_1, U_2)$. Lemma 3 gives a tighter bound in many cases. Consider $(U_1, U_2)$ with the joint distribution: 
$P(U_1=0;U_2=0)= P(U_1=1;U_2=1)=0.4444;  P(U_1=1;U_2=0)= P(U_1=0;U_2=1)=0.0556$. The correlation between the sources is 0.7778 but from Lemma 3, the correlation between $(X_1, X_2)$ cannot exceed 0.7055.
\subsection {A coding Scheme}\label{sec1.5.1}

In this section we develop a distributed coding scheme for mapping the discrete alphabets $(U_1,U_2)$ into jointly Gaussian correlated code words $(X_1,X_2)$ which  satisfy \eqref{covereq} and the Markov condition. The heart of the scheme is to approximate a jointly Gaussian distribution with the sum of product of Gaussian marginals. Although this is stated in the following lemma for two dimensional vectors $(X_1, X_2)$, the results hold for any finite dimensional vectors (hence can be used for any number of users sharing the MAC).

\begin{lemma}
Any jointly Gaussian two dimensional density can be uniformly arbitrarily closely approximated by a weighted sum of product of marginal Gaussian densities:
\begin{gather}
\sum_{i=1}^N{\frac{p_i}{\sqrt{2\pi c_{1i}}}e^{\frac{-1}{2c_{1i}}(x_1-a_{1i})^2}\frac{q_i}{\sqrt{2\pi c_{2i}}}e^{\frac{-1}{2c_{2i}}(x_2-a_{2i})^2}}
\label{co2} 
\end{gather}
\label{lemma4}
\end{lemma}

$Proof$: See Appendix~\ref{l1}.~~~~~~~~~~~~~~~~~~~~~~~~~~~~~~~~~~~~~~~~~~~~~~~~~~~~~~~~~~~~~~~~~~~~~~~~~~~~~~~~~~~~~~~~~~~~~~~~~~~~~~~~~~~~~~~~$\blacksquare$ 

From the above lemma we can form a sequence of functions $f_n(x_1,x_2)$ of type \eqref{co2} such that $sup_{x_1,x_2}|f_n(x_1,x_2)-f(x_1,x_2)| \rightarrow 0$ as $ n \rightarrow \infty$, where $f$ is a given jointly Gaussian density. Although $f_n$ are not guaranteed to be probability densities, due to uniform convergence, for large $n$, they will almost be. In the following lemma we will assume that we have made the minor modification to ensure that $f_n$ is a proper density for large enough $n$. This lemma shows that obtaining $(X_1, X_2)$ from such approximations can provide the (relaxed) upper bounds in \eqref{sub1}-\eqref{sub3} (we actually show for the third inequality only but this can be shown for the other inequalities in the same way). Of course, as mentioned earlier, then these can be used to obtain the $(X_1,X_2)$ which satisfy the actual bounds in \eqref{covereq}.

Let $(X_{m1},X_{m2})$  and $(X_1, X_2)$ be random variables with densities $f_m$ and $f$ and $sup_{x_1,x_2}|f_m(x_1,x_2)-f(x_1,x_2)|\rightarrow0$ as $ m \rightarrow \infty$. Let $Y_m$ and $Y$ denote the corresponding channel outputs. 

\begin{lemma}
 For the random variables defined above, if $\{logf_m(Y_m),m\geq1\}$  is uniformly integrable, $I(X_{m1},X_{m2};Y_m) \rightarrow I(X_1,X_2;Y)$ as $m \rightarrow \infty$.
\label{lemma5}
\end{lemma}

$Proof$: See Appendix~\ref{l1}.~~~~~~~~~~~~~~~~~~~~~~~~~~~~~~~~~~~~~~~~~~~~~~~~~~~~~~~~~~~~~~~~~~~~~~~~~~~~~~~~~~~~~~~~~~~~~~~~~~~~~~~~~~~~~~~~$\blacksquare$  

A set of sufficient conditions for uniform integrability of $\{logf_m(Y_m),m\geq1\}$  is 

(1) Number of components in \eqref{co2} is upper bounded.

(2) Variance of component densities in \eqref{co2} is upper bounded and lower bounded away from zero.

(3) The means of the component densities in \eqref{co2} are in a bounded set.

From Lemma \ref{lemma4} a joint Gaussian density with any correlation  can be expressed by a linear combination of marginal Gaussian densities. But the coefficients $p_i$ and $q_i$ in \eqref{co2} may be positive or negative. To realize our coding scheme, we would like to have the $p_i$'s and $q_i$'s to be non negative. This introduces constraints on the realizable Gaussian densities in our coding scheme. For example, from Lemma \ref{lemma3}, the correlation $\rho$  between $X_1$ and $X_2$ cannot exceed $\sqrt{1-2^{-2I(U_1;U_2)}}$. Also there is still the question of getting a good linear combination of marginal densities to obtain the joint density for a given $N$ in \eqref{co2}.
 	
This motivates us to consider an optimization procedure for finding $p_i,~q_i$, $a_{1i},~a_{2i}$, $c_{1i}$ and $c_{2i}$ in \eqref{co2} that provides the best approximation to a given joint Gaussian density. We illustrate this with an example. Consider $U_1, U_2$ to be binary. Let $ P(U_1=0; U_2=0)=p_{00}; P(U_1=0; U_2=1)=p_{01};  P(U_1=1; U_2=0)=p_{10} $ and $ P(U_1=1; U_2=1)= p_{11}$. Define (notation in the following has been slightly changed compared to \eqref{co2})
\begin{gather}
f(X_1=.|U_1=0)=p_{101}\mathcal{N}(a_{101},c_{101})+p_{102}\mathcal{N}(a_{102},c_{102})\nonumber\\
...+p_{10r_1}\mathcal{N}(a_{10r_1},c_{10r_1}),\label{newlab1}\\ 
f(X_1=.|U_1=1)=p_{111}\mathcal{N}(a_{111},c_{111})+p_{112}\mathcal{N}(a_{112},c_{112})\nonumber\\
...+p_{11r_2}\mathcal{N}(a_{11r_2},c_{11r_2}),\\
f(X_2=.|U_2=0)=p_{201}\mathcal{N}(a_{201},c_{201})+p_{202}\mathcal{N}(a_{202},c_{202})\nonumber\\
...+p_{20r_3}\mathcal{N}(a_{20r_3},c_{20r_3}),\\
f(X_2=.|U_2=1)=p_{211}\mathcal{N}(a_{211},c_{211})+p_{212}\mathcal{N}(a_{212},c_{212})\nonumber\\
...+p_{21r_4}\mathcal{N}(a_{21r_4},c_{21r_4})\label{newlab4}.
\end{gather} 
where $\mathcal{N}(a,b)$  denotes Gaussian density with mean $a$  and variance $b$. Let $\underline{p}$  be the vector with components $p_{101},...,p_{10r_1}$, $p_{111},...,p_{11r_2}$, $p_{201},...,p_{20r_3}$, $p_{211},...,p_{21r_4}$. Similarly we denote by $\underline{a}$  and $\underline{c}$ the vectors with components $a_{101},...,a_{10r_1}$, $a_{111},...,a_{11r_2}$, $a_{201},...,a_{20r_3}$, $a_{211},...,a_{21r_4}$ and $c_{101},...,c_{10r_1}$, $c_{111},...,c_{11r_2}$, $c_{201},...$,$c_{20r_3}$, $c_{211}$,$...,c_{21r_4}$. The mixture of Gaussian densities \eqref{newlab1}-\eqref{newlab4} will be used to obtain the RHS in \eqref{co2} for an optimal approximation. For a given $\underline{p},~\underline{a},~\underline{c}$, the resulting joint density is  $g_{\underline{p},\underline{a},\underline{c}} = p_{00}f(X_1=.|U_1=0)f(X_2=.|U_2=0)+p_{01}f(X_1=.|U_1=0)f(X_2=.|U_2=1)+p_{10}f(X_1=.|U_1=1)f(X_2=.|U_2=0)+p_{11}f(X_1=.|U_1=1)f(X_2=.|U_2=1)$.

Let $f_\rho(x_1,x_2)$ be the jointly Gaussian density that we want to approximate. Let it has zero mean and covariance matrix $K_{X_1,X_2}=\begin{pmatrix}
1 & \rho \\
\rho &  1
\end{pmatrix}$.  The best $g_{\underline{p},\underline{a},\underline{c}}$ is obtained by solving the minimization problem:
\begin{gather}
\text{min}_{\underline{p},\underline{a},\underline{c}}\int{[g_{\underline{p},\underline{a},\underline{c}}(x_1,x_2)- f_\rho(x_1,x_2)]^2 dx_1dx_2}
\label{optim}
\end{gather}
subject to 
\begin{gather}
(p_{00}+p_{01})\sum_{i=1}^{r_1}p_{10i}a_{10i}+(p_{10}+p_{11})\sum_{i=1}^{r_2}p_{11i}a_{11i}=0,\nonumber
\end{gather}
\begin{gather}
(p_{00}+p_{10})\sum_{i=1}^{r_3}p_{20i}a_{20i}+(p_{01}+p_{11})\sum_{i=1}^{r_4}p_{21i}a_{21i}=0,\nonumber
\end{gather}
\begin{gather}
(p_{00}+p_{01})\sum_{i=1}^{r_1}p_{10i}(c_{10i}+a_{10i}^2)+\nonumber (p_{10}+p_{11})\sum_{i=1}^{r_2}p_{11i}(c_{11i}+a_{11i}^2)=1,\nonumber
\end{gather}
\begin{gather}
(p_{00}+p_{10})\sum_{i=1}^{r_3}p_{20i}(c_{20i}+a_{20i}^2)+\nonumber
(p_{01}+p_{11})\sum_{i=1}^{r_4}p_{21i}(c_{21i}+a_{21i}^2)=1,\nonumber
\end{gather}
\begin{gather}
\sum_{i=1}^{r_1}p_{10i}=1,\sum_{i=1}^{r_2}p_{11i}=1,\sum_{i=1}^{r_3}p_{20i}=1,\sum_{i=1}^{r_4}p_{21i}=1,\nonumber
\end{gather}
\begin{gather}
p_{10i}\geq 0, c_{10i}\geq 0 ~for~i \in \{1,2...r_1\},~\nonumber 
p_{11i}\geq 0, c_{11i}\geq 0 ~for~ i \in \{1,2...r_2\},\nonumber \\
p_{20i}\geq 0, c_{20i}\geq 0~ for~i \in \{1,2...r_3\},~\nonumber 
p_{21i}\geq 0, c_{21i}\geq 0 ~for~ i \in \{1,2...r_4\}.\nonumber 
\label{const}
\end{gather}
The above constraints are such that the resulting distribution $g$ for $(X_1,X_2)$ will satisfy $E[X_i]=0$ and $E[X_i^2]=1,~i=1,2$.

The above coding scheme will be used to obtain a codebook as follows. If user 1 produces $U_1=0$, then independently with probability $p_{10i}$
  the encoder 1 obtains codeword $X_1$   from the distribution $\mathcal{N}(a_{10i},c_{10i})$ independently of other codewords. Similarly we obtain the codewords for $U_1=1$ and for user 2. Once we have found the encoder maps the encoding and decoding are as described in the proof of Theorem 1. The decoding is done  by joint typicality of the received $Y^n$ with $(U_1^n,U_2^n)$. 

This coding scheme can be extended to any discrete alphabet case. We give an example below to illustrate the coding scheme.
\subsection{Example }
Consider $(U_1, U_2)$ with the joint distribution: 
$P(U_1=0; U_2=0) = P(U_1=1; U_2=1)= P(U_1=0; U_2=1)=1/3;P(U_1=1; U_2=0)=0$
and power constraints $P_1 = 3 ; P_2 = 4$. Also consider a GMAC with $\sigma_N^2 =1$. If the sources are mapped into independent channel code words, then the sum rate condition in \eqref{sub3} with $\rho=0$  should hold. The LHS evaluates to 1.585 bits whereas the RHS is 1.5 bits. Thus  \eqref{sub3} is violated and hence the sufficient conditions in \eqref{covereq} are also violated. 

In the following we explore the possibility of using correlated $(X_1, X_2)$ to see if we can transmit this source on the given MAC. The inputs $(U_1, U_2)$ can be distributedly mapped to jointly Gaussian channel code words $(X_1, X_2)$ by the technique mentioned above. The maximum $\rho$ which satisfies upper bounds in  \eqref{sub1} and \eqref{sub2} are 0.7024 and 0.7874 respectively and the minimum $\rho$  which satisfies \eqref{sub3} is 0.144.  From Lemma~\ref{lemma3}, $\rho$ is upper bounded by 0.546. Therefore we want to obtain jointly Gaussian $(X_1, X_2)$ satisfying $X_1\leftrightarrow U_1\leftrightarrow U_2\leftrightarrow X_2$ with correlation $\rho \in [0.144,0.546]$.  If we choose $\rho=0.3$, it meets the inner bounds in  \eqref{sub1}-\eqref{sub3} (i.e., the bounds in \eqref{covereq}): $I (X_1;Y | X_2, U_2) =0.792,~I (X_2;Y | X_1, U_1)=0.996$, $I ( X_1;Y | X_2 )=0.949,~ I ( X_2;Y | X_1)=1.107$,  $H(U_1|U_2)=H(U_2|U_1)=0.66$.

We choose $r_i = 2,~i=1,...,4$ and  solve the  optimization problem \eqref{optim} via MATLAB to get the function $g$. The optimal solution solution has both component distributions in \eqref{newlab1}- \eqref{newlab4} same and these are
\begin{gather}
f(X_1|U_1=0)=\mathcal{N}(-0.0002,0.9108),~\nonumber
f(X_1|U_1=1)=\mathcal{N}(-0.0001,1.0446),\nonumber\\
f(X_2|U_2=0)=\mathcal{N}(-0.0021,1.1358),~\nonumber
f(X_2|U_2=1)=\mathcal{N}(-0.0042,0.7283).\nonumber
\label{resul}
\end{gather}
The normalized minimum distortion, defined as ${\int{[g_{\underline{p},\underline{a},\underline{c}}(x_1,x_2)- f_\rho(x_1,x_2)]^2 dx_1dx_2}}/$ ${\int{f_\rho^2(x_1,x_2)dx_1dx_2}}$ is 0.137\%.

\begin{figure}[h]
\centering
\includegraphics [width=2.4in,height=2in] {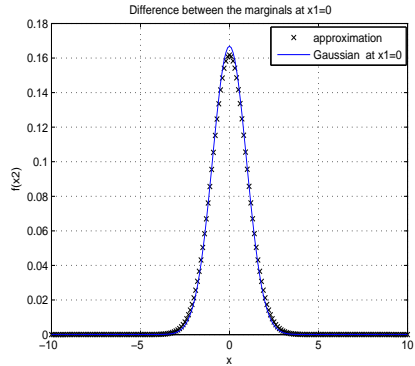}
\caption{Cross section of the approximation of the joint Gaussian with $\rho$=0.3.}
\label{figapprox1}
\end{figure}

The approximation (a cross section of the two dimensional densities) is shown in Fig.~\ref{figapprox1}.

\begin{figure}[h]
\centering
\includegraphics [width=2.4in,height=2in] {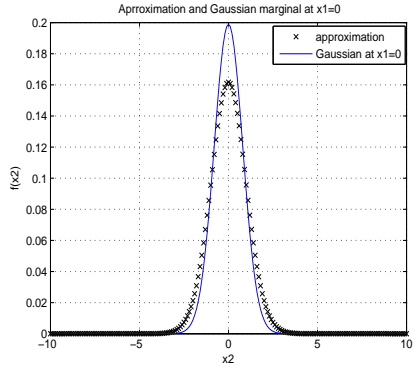}
\caption{Cross section of the approximation of the joint Gaussian with $\rho$=0.6.}
\label{figapprox2}
\end{figure}

If we take $\rho =0.6$ which violates Lemma~\ref{lemma3} then the optimal solution from \eqref{optim} is shown in Fig.~\ref{figapprox2}.  We can see that the error in this case is more. Now the normalized marginal distortion is 10.5 \%.

\subsection{Generalizations}
The procedure mentioned in Section \ref{sec1.5.1} can be extended to systems with general discrete alphabets, multiple sources, lossy transmissions and side information as follows.

Consider  $N \ge 2$ users with  source $i$ taking values in a discrete alphabet $\mathcal{U}_i$.  In such a case for each user we find $P(X_i=.|{U_i=u_i}),~u_i \in \mathcal{U}_i$ using a mapping mentioned as in \eqref{newlab1}-\eqref{newlab4} to yield jointly Gaussian $(X_1,X_2,...,X_N)$.

If $Z_1$ and $Z_2$ are the side information available, then we use $f(X_i=.|U_i,Z_i),~ i={1,2}$ as in  \eqref{newlab1}-\eqref{newlab4} and obtain the optimal approximation from \eqref{optim}.

For lossy transmission, we choose appropriate  discrete auxiliary random variables $W_i$  satisfying the conditions in  Theorem 1. Then we can form $(X_1, X_2)$ from $(W_1, W_2)$ via the optimization procedure \eqref{optim}.

\section{Gaussian sources over a GMAC}\label{sec1.6}
In this section we consider transmission of correlated Gaussian sources over a GMAC. This is an important example for transmitting continuous alphabet sources over a GMAC. For example one comes across it if a sensor network is sampling a Gaussian random field. Also, in the application of detection of change (\cite{Veeravalli01decentralised}) by a sensor network, it is often the detection of change in the mean of the sensor observations with the sensor observation noise being Gaussian.

We will  assume that $(U_{1n}, U_{2n})$ is jointly Gaussian with mean zero, variances $\sigma_i^2,~i=1,2$   and correlation $ \rho. $ The distortion measure will be Mean Square Error (MSE). The (relaxed) sufficient conditions from \eqref{sub1}-\eqref{sub3} for transmission of the sources over the channel are given by (these continue to hold because Lemmas~\ref{lemma1}-\ref{lemma3} are still valid)
\begin{eqnarray}
I (U_1; W_1 | W_2)  <  0.5\log\left[ 1+ \frac{P_1(1- {\tilde{\rho}}^{2})} {{\sigma_N}^{2}}\right],~
I (U_2; W_2 | W_1)   <   0.5\log\left[ 1+ \frac{P_2(1- {\tilde{\rho}}^{2})} {{\sigma_N}^{2}}\right],\label{gogmac}\\   			        
I (U_1, U_2 ; W_1, W_2)   <    0.5\log\left[ 1+ \frac{P_1 + P_2 + {2} {\tilde{\rho}}{\sqrt{P_1P_2}}} {{\sigma_N}^{2}} \right]\nonumber.~~~~~~~~~~~~~~~~~~
\end{eqnarray}
where $\tilde{\rho}$  is the correlation between $(X_1,X_2)$ which are chosen to be jointly Gaussian, as in Section \ref{sec1.5}.

We consider three specific coding schemes to obtain $W_1,W_2,X_1,X_2$  where $(W_1,W_2)$   satisfy the distortion constraints and $(X_1,X_2)$ are jointly Gaussian with an appropriate $\tilde{\rho}$  such that \eqref{gogmac} is satisfied. These coding schemes have been widely used. The schemes are Amplify and Forward (AF), Separation Based (SB) and the coding scheme provided in Lapidoth and Tinguely (LT) \cite{Lapidoth01sending}.  We have compared the performance of these schemes in \cite{Rajesh07allerton}. The AF and LT are joint source-channel coding schemes. In \cite{Lapidoth01sending} it is shown that AF is optimal at low SNR. In \cite{Rajesh07allerton} we show that at high SNR LT is close to optimal. SB although performs well at high SNR,  is sub-optimal.

For general continuous alphabet sources $(U_1,U_2)$, no necessarly Gaussian, we vector quantize $U_1^n,U_2^n$ into $\tilde{U}_1^n,\tilde{U}_2^n$. Then to obtain correlated Gaussian codewords $(X_1^n,X_2^n)$ we can use the scheme provided in Section \ref{sec1.5.1}. Alternatively, use Slepian-Wolf coding on  $(\tilde{U}_1^n,\tilde{U}_2^n)$. Then for large $n$, $\tilde{U}_1^n$ and $\tilde{U}_2^n$ are almost independent. Now on each $\tilde{U}_i^n,~i=1,2$ we can use usual independent Gaussian codebooks as in a point to point channel.

\section{Conclusions}\label{sec1.12}
In this paper, sufficient conditions are provided for transmission of correlated sources over a multiple access channel. Various previous results on this problem are obtained as special cases.  Suitable examples are given to emphasis the superiority of joint source-channel coding schemes. Important special cases of  correlated discrete sources over a GMAC and Gaussian sources over a GMAC are discussed in more detail. In particular a new joint source-channel coding scheme is presented for discrete sources over a GMAC.

\appendices
\section{Proof of Theorem 1}
The coding scheme involves distributed quantization $(W_1^n,W_2^n)$ of the sources and the side information $(U_1^n,Z_1^n),(U_2^n,Z_2^n)$  followed by a correlation preserving mapping to the channel codewords. The decoding approach involves first decoding  $(W_1^n,W_2^n)$  and then obtaining estimate $(\hat{U}_1^n,\hat{U}_2^n)$ as a function of $(W_1^n,W_2^n)$ and the decoder side information $Z^n$. 

Let $T_\epsilon^n(X,Y)$ denote the weakly $\epsilon$-typical set of sequences of length $n$ for $(X,Y)$ where $\epsilon >0$ is  an arbitrarily small fixed positive constant.
We  use the following  Lemmas in the proof.

{\bf{Markov Lemma}}: Suppose $X\leftrightarrow Y \leftrightarrow Z$. If for a given $(x^n,y^n) \in T_\epsilon^n(X,Y)$, $Z^n$ is drawn according to $\prod_{i=1}^np(z_i|y_i)$, then with high probability $(x^n,y^n,Z^n) \in T_\epsilon^n(X,Y,Z)$ for $n$ sufficiently large.
\label {ls}

The proof of this Lemma for strong typicality is given in \cite{tb}. We need it for weak typicality. By the Markov property, $(x^n,y^n,z^n)$ formed in the statement of the Lemma has the same joint distribution as the original sequence $(X^n,Y^n,Z^n)$. Thus the statement of the above Lemma follows. In the same way the following Lemma also holds.

{\bf{Extended Markov Lemma}}: Suppose $W_1\leftrightarrow U_1Z_1 \leftrightarrow U_2W_2Z_2Z$ and  $W_2\leftrightarrow U_2Z_2 \leftrightarrow U_1W_1Z_1Z$. If for a given  $(u_1^n,u_2^n,z_1^n,z_2^n,z^n)$  $\in$    $T_\epsilon^n(U_1,U_2,Z_1,Z_2,Z)$, $W_1^n$ and $W_2^n$ are drawn respectively according to  $\prod_{i=1}^np(w_{1i}|u_{1i},z_{1i})$ and  $\prod_{i=1}^np(w_{2i}|u_{2i},z_{2i})$, then with high probability  $(u_1^n,u_2^n,z_1^n,z_2^n$,$z^n,W_1^n$,$W_2^n)$ $\in$  $T_\epsilon^n(U_1,U_2,Z_1,Z_2,Z,W_1,W_2)$ for $n$ sufficiently large.


We show the achievability of all points in the rate region (1).

$Proof$:  Fix $p(w_1|u_1,z_1), p(w_2|u_2,z_2),p(x_1|w_1),p(x_2|w_2)$ as well as $f_D{n}(.)$ satisfying the distortion constraints. First we give the proof for the discrete channel alphabet case.

$Codebook~ Generation$: Let $R_i^{'}=I(U_i,Z_i;W_i)+\delta, i \in \{1,2\}$ for some $\delta >0$. Generate $2^{nR_i^{'}}$ codewords of length $n$, sampled iid from the marginal distribution $p(w_i), i \in \{1,2\}$. For each $w_i^n$ independently generate sequence $X_i^n$ according to $\prod_{j=1}^n p(x_{ij}|w_{ij}), i \in \{1,2\}$. Call these sequences $x_i(w_i^n), i \in {1,2}$. Reveal the codebooks to the encoders and the decoder.

$Encoding$: For $i \in \{1,2\}$, given the source sequence $U_i^n$ and  $Z_i^n$, the $i^{th}$ encoder looks for a codeword $W_i^n$ such that $(U_i^n,Z_i^n,W_i^n) \in T_{\epsilon}^n(U_i,Z_i,W_i)$ and then transmits $X_i(W_i^n)$. 

$Decoding$: Upon receiving $Y^n$, the decoder finds the unique $(W_1^n,W_2^n)$ pair such that\\ $(W_1^n,W_2^n,x_1(W_1^n),x_2(W_2^n),Y^n,Z^n)\in T_{\epsilon}^n$. If it fails to find such a unique pair, the decoder declares an error and incurres a maximum distortion of $d_{max}$ (we assume that the distortion measures are bounded; at the end we will remove this condition).

In the following we show that the probability of error for this encoding-decoding scheme tends to zero as $n \rightarrow \infty$. The error can occur because of the following four events {\bf{E1}}-{\bf{E4}}. We show that $P({\bf{Ei}}) \rightarrow 0$, for ${\bf{i}}= 1,2,3,4$.

{\bf{E1}} The encoders do not find the codewords. However from rate distortion theory (\cite{Cover04elements}, page 356), $\lim_{n \to \infty}P(E_1)=0$ if $R_i^{'} > I (U_i,Z_i;W_i), i \in {1,2}$.

{\bf{E2}} The codewords are not jointly typical with $(Y^n,Z^n)$.
Probability of this event goes to zero from the extended Markov Lemma.

{\bf{E3}} There exists another codeword $\hat{w}_1^n$ such that $(\hat{w}_1^n,W_2^n,x_1(\hat{w}_1^n),x_2(W_2^n)$,\\ $Y^n,Z^n)\in T_\epsilon^n$.  Define $ \alpha {\buildrel\Delta \over=}$  $(\hat{w}_1^n,W_2^n,x_1(\hat{w}_1^n),x_2(W_2^n),Y^n,Z^n)$. Then,
\begin{eqnarray}
 P({\bf{E3}}) = Pr {\{\text{There is} ~ \hat{w}_1^n \ne w_1^n : \alpha \in T_{\epsilon}^n}\}
 \leq \sum_{\hat{w}_1^n \ne W_1^n:(\hat{w}_1^n,W_2^n,Z^n) \in T_{\epsilon} ^n} Pr{\{\alpha \in T_{\epsilon}^n\}}
 \label{e1}
\end{eqnarray}

The  probability term inside the summation in (\ref{e1}) is 
\begin{eqnarray*}
&\leq& \sum_{(x_1(.),x_2(.),y^n):{\alpha \in T_{\epsilon}^n}} Pr\{x_1(\hat{w}_1^n),x_2(w_2^n),y^n|\hat{w}_1^n,w_2^n,z^n\} p(\hat{w}_1^n,w_2^n,z^n)\\
&\leq& \sum_{(x_1(.),x_2(.),y^n):{\alpha \in T_{\epsilon}^n}} Pr\{x_1(\hat{w}_1^n)|\hat{w}_1^n\}Pr\{x_2(w_2^n),y^n|w_2^n,z^n\}\\
&\leq& \sum_{(x_1(.),x_2(.),y^n):{\alpha \in T_{\epsilon}^n}} 2^{-n\{H(X_1|W_1)+H(X_2,Y|W_2,Z)-4\epsilon\}}\\
&\leq& 2^{n{H(X_1,X_2,Y|W_1,W_2,Z)}}2^{-n\{H(X_1|W_1)+H(X_2,Y|W_2,Z)-4\epsilon\}} \nonumber.
\label{e2}
\end{eqnarray*}
But from hypothesis, we have
\begin{eqnarray*}
&&H(X_1,X_2,Y|W_1,W_2,Z)- H(X_1|W_1)-H(X_2,Y|W_2,Z) \\
&=&H(X_1|W_1)+H(X_2|W_2)+H(Y|X_1,X_2)-H(X_1|W_1)-H(X_2,Y|W_2,Z)\\
&=&H(Y|X_1,X_2)-H(Y|X_2,W_2,Z)\\
&=&H(Y|X_1,X_2,W_2,Z)-H(Y|X_2,W_2,Z) =~ -I(X_1;Y|X_2,W_2,Z).
\label{e3}
\end{eqnarray*}
Hence,
\begin{gather}
Pr\{(\hat{w}_1^n,W_2^n,x_1(\hat{w}_1^n),x_2(W_2^n),Y^n,Z^n)\in T_{\epsilon}^n\}
\leq 2^{-n\{I(X_1;Y|X_2,W_2,Z)-6\epsilon\}}.
\label{e4}
\end{gather}
Then from \eqref{e1}
\begin{eqnarray}
P({\bf{E3}})&\leq& \sum_{\hat{w}_1^n \ne w_1^n:(\hat{w}_1^n,w_2^n,z^n)\in T_{\epsilon}^n} 2^{-n\{I(X_1;Y|X_2,W_2,Z)-6\epsilon\}}\nonumber\\
 &=& |\{\hat{w}_1^n:(\hat{w}_1^n,w_2^n,z^n)\in T_{\epsilon}^n\}|2^{-n\{I(X_1;Y|X_2,W_2,Z)-6\epsilon\}}\nonumber\\
 &\leq&|\{\hat{w}_1^n\}|Pr\{\hat{w}_1^n,w_2^n,z^n)\in T_{\epsilon}^n\}2^{-n\{I(X_1;Y|X_2,W_2,Z)-6\epsilon\}}\nonumber\\
 &\leq& 2^{n\{I(U_1,Z_1;W_1)+\delta\}}2^{-n\{I(W_1;W_2,Z)-3\epsilon\}}
 2^{-n\{I(X_1;Y|X_2,W_2,Z)-6\epsilon\}}\label{added}\\
 &=&2^{n\{I(U_1,Z_1;W_1|W_2,Z)\}}2^{-n\{I(X_1;Y|X_2,W_2,Z)-9\epsilon-\delta\}}.\nonumber
\end{eqnarray}
In \eqref{added} we have used the fact that
\begin{eqnarray*}
I(U_1,Z_1;W_1)-I(W_1;W_2,Z)=H(W_1|W_2,Z)-H(W_1|U_1,Z_1)\\
=H(W_1|W_2,Z)-H(W_1|U_1,Z_1,W_2,Z)=I(U_1,Z_1;W_1|W_2,Z).
\label{e6}
\end{eqnarray*}
The RHS of \eqref{added} tends to zero if $ I(U_1,Z_1;W_1|W_2,Z) < I (X_1;Y|X_2,W_2,Z)$.

Similarly, by symmetry of the problem we require 
$ I(U_2,Z_2;W_2|W_1,Z) < I (X_2;Y|X_1,W_1,Z)$.

{\bf{E4}} There exist other codewords $\hat{w}_1^n$ and $\hat{w}_2^n$  such that $\alpha {\buildrel\Delta\over =}(\hat{w}_1^n,\hat{w}_2^n,x_1(\hat{w}_1^n),x_2(\hat{w}_2^n),Y^n,Z^n)\in T_{\epsilon}^n$. Then,
\begin{eqnarray}
 P({\bf{E4}}) &=& Pr {\{\text{There is} ~ (\hat{w}_1^n,\hat{w}_2^n)  \ne (w_1^n,w_2^n) : \alpha \in T_{\epsilon}^n}\}\nonumber \\
 &\leq& \sum_{(\hat{w}_1^n,\hat{w}_2^n) \ne (w_1^n,w_1^n):(\hat{w}_1^n,\hat{w}_2^n,z^n) \in T_{\epsilon}^n} Pr{\{\alpha \in T_{\epsilon}^n\}}.
 \label{e10}
\end{eqnarray}

The  probability term inside the summation in (\ref{e10}) is 
\begin{eqnarray*}
&\leq& \sum_{(x_1(.),x_2(.),y^n):{\alpha \in T_{\epsilon}^n}} Pr\{x_1(\hat{w}_1^n),x_2(\hat{w}_2^n),y^n|\hat{w}_1^n,\hat{w}_2^n,z^n\}p(\hat{w}_1^n,\hat{w}_2^n,z^n)\\
&\leq& \sum_{....}{Pr\{x_1(\hat{w}_1^n)|\hat{w}_1^n\}Pr\{x_2(\hat{w}_2^n)|\hat{w}_2^n\} Pr\{y^n|z^n\}}\\
&\leq& \sum_{(x_1(.),x_2(.),y^n):{\alpha \in T_{\epsilon}^n}} 2^{-n\{H(X_1|W_1)+H(X_2|W_2)+H(Y|Z)-5\epsilon\}}\\
&\leq& 2^{n{H(X_1,X_2,Y|W_1,W_2,Z)}} 2^{-n\{H(X_1|W_1)+H(X_2|W_2)+H(Y|Z)-7\epsilon\}}.\nonumber
\label{e11}
\end{eqnarray*}
But from hypothesis, we have
\begin{eqnarray*}
H(X_1,X_2,Y|W_1,W_2,Z)- H(X_1|W_1)-H(X_2|W_2)-H(Y|Z)~~~~~~~~~~~~~~~~\\
=H(Y|X_1,X_2)-H(Y|Z)=H(Y|X_1,X_2,Z)-H(Y|Z)=-I(X_1,X_2;Y|Z).
\label{e12}
\end{eqnarray*}
Hence,
\begin{gather}
Pr\{(\hat{w}_1^n,\hat{w}_2^n,x_1(\hat{w}_1^n),x_2(\hat{w}_2^n),y^n,z^n)\in T_{\epsilon}^n\}
\leq 2^{-n\{I(X_1,X_2;Y|Z)-7\epsilon\}}.
\label{e13}
\end{gather}
Then from \eqref{e10}
\begin{eqnarray*}
P({\bf{E4}}) &\leq& \sum_{\substack{(\hat{w}_1^n,\hat{w}_2^n) \ne (w_1^n,w_1^n):\\(\hat{w}_1^n,\hat{w}_2^n,z^n) \in T_{\epsilon}^n}} 2^{-n\{I(X_1,X_2;Y|Z)-7\epsilon\}}\\
 &=& |\{(\hat{w}_1^n,\hat{w}_2^n):(\hat{w}_1^n,\hat{w}_2^n,z^n)\in T_{\epsilon}^n\}|2^{-n\{I(X_1,X_2;Y|Z)-7\epsilon\}}\\
 &\leq&|\{\hat{w}_1^n\}||\{\hat{w}_2^n\}|Pr\{(\hat{w}_1^n,\hat{w}_2^n,z^n)\in T_{\epsilon}^n\}
 2^{-n\{I(X_1,X_2;Y|Z)-7\epsilon\}}\nonumber\\
&\leq& 2^{n\{I(U_1,Z_1;W_1)+I(U_2,Z_2;W_2)+2\delta\}}.\\
&&2^{-n\{I(W_1;W_2,Z)+I(W_2;W_1,Z)+I(W_1;W_2|Z)-4\epsilon\}}
 2^{-n\{I(X_1,X_2;Y|Z)-7\epsilon\}}\nonumber\\
 &=&2^{n\{I(U_1,U_2,Z_1,Z_2;W_1,W_2|Z)\}}2^{-n\{I(X_1,X_2;Y|Z)-11\epsilon-2\delta\}}.
\label{e15}
\end{eqnarray*}
The RHS of the above inequality tends to zero if $ I(U_1,U_2,Z_1,Z_2;W_1W_2|Z) < I(X_1,X_2;Y|Z)$.

Thus as $n \rightarrow \infty$, with probability tending to 1, the decoder finds the correct sequence $(W_1^n,W_2^n)$ which is jointly  weakly $\epsilon$-typical with $(U_1^n,U_2^n,Z^n)$.

The fact that $(W_1^n,W_2^n)$ is weakly $\epsilon$-typical with $(U_1^n,U_2^n,Z^n)$ does not guarantee that $f_D^n(W_1^n,W_2^n,Z^n)$ will satisfy the distortions $D_1,D_2$. For this, one needs that $(W_1^n,W_2^n)$ is distortion-$\epsilon$-weakly typical (\cite{Cover04elements}) with   $(U_1^n,U_2^n,Z^n)$. Let $T_{D,\epsilon}^n$  denote the set of distortion typical sequences. Then by strong law of large numbers  $P(T_{D,\epsilon}^n|T_\epsilon^n)\rightarrow 1$ as $n\rightarrow \infty$. Thus the distortion constraints are also satisfied by $(W_1^n,W_2^n)$ obtained above with a probability tending to 1 as $n \rightarrow \infty$. Therefore, if distortion measure $d$ is bounded 
$\lim_{n \rightarrow \infty}E[d(U_i^n,\hat{U}_i^n)] \leq D_i+\epsilon,~i=1,2$.

For continuous channel alphabet case (e.g., GMAC) one also needs transmission constraints $E[g_i(X_i)] \le \alpha_i,~i=1,2$.  For this we need to ensure that the coding scheme chooses a distribution $p(x_i|w_i)$ which satisfies $E[g_i(X_i)] < \alpha_i-\epsilon$. Then if a specific codeword does not satisfy $\frac{1}{n} \sum_{k=1}^n g_i(x_k) < \alpha_i$, one declares an error. As $ n \rightarrow \infty$ this happens with a vanishingly small probability.

If there exist $u_i^*$ such that $E[d_i(U_i,u_i^*)]<\infty,~i = 1,2 $, then the result extends to unbounded distortion measures also as follows. Whenever the decoded $(W_1^n,W_2^n)$ are not in the distortion typical set then we estimate $(\hat{U}_1^n,\hat{U}_2^n)$ as $({u_1^*}^n,{u_2^*}^n)$. Then for $i=1,2$,
\begin{equation}
\label{ala}
E[d_i(U_i^n,\hat{U}_i^n)] \leq D_i+\epsilon + E[d(U_i^n,{u_i^*}^n) {\bf{1}}_{\{(T_{D,\epsilon}^n)^c\}}].
\end{equation}
Since $E[d(U_i^n,{u_i^*}^n)] < \infty $ and $P[({T_{D,\epsilon}^n})^c] \rightarrow 0$ as $n \rightarrow \infty$, the last term of \eqref{ala} goes to zero as  $n \rightarrow \infty$.
\section{ Proof of converse for lossless transmission of discrete correlated sources over orthogonal channels with side information}
Let $P^e_n$ be the probability of error in estimating $U_1^n,~U_2^n$ from $(Y_1^n,Y_2^n,Z^n)$. For any given coding-decoding scheme, we will show that if $P^e_n \rightarrow 0$ then the inequalities in \eqref{conints2} specialized to the lossless transmission must be satisfied for this system. 

Let $||\mathcal{U}_i||$ be the cardinality of set $\mathcal{U}_i$. From Fano's inequality we have
\begin{eqnarray*}
\frac{1}{n} H(U_1^n,U_2^n|Y_1^n,Y_2^n,Z^n) &\le&  \frac{1}{n} log (||U_1^n  U_2^n||)P^e_n + \frac{1}{n}\\
&=& P^e_n(log ||\mathcal{U}_1||+log||\mathcal{U}_2||) + \frac{1}{n}.
\end{eqnarray*}
Denote $P_n(log ||\mathcal{U}_1||+log||\mathcal{U}_2||) + \frac{1}{n}$ by $\lambda_n$. As $P^e_n \rightarrow 0, \lambda_n \rightarrow 0$.

Since,
\begin{eqnarray*}
H(U_1^n,U_2^n|Y_1^n,Y_2^n,Z^n)= H(U_1^n|Y_1^n,Y_2^n,Z^n)+ H(U_2^n|U_1^n,Y_1^n,Y_2^n,Z^n),
\end{eqnarray*}
we obtain $H(U_1^n|Y_1^n,Y_2^n,Z^n)/n \le \lambda_n$. Therefore, because $\{U_1^n\}$ is an iid sequence,
\begin{eqnarray}
n H(U_1)&=&H(U_1^n)\nonumber\\
&=& H(U_1^n|Y_1^n,Y_2^n,Z^n)+ I (U_1^n;Y_1^n,Y_2^n,Z^n)\nonumber\\
&\le& n \lambda_n+ I(U_1^n; Y_1^n,Y_2^n,Z^n)\label{ap1}.
\end{eqnarray}
Also, by data processing inequality,
\begin{eqnarray}
I(U_1^n; Y_1^n,U_2^n,Z^n)&=& I(U_1^n;U_2^n,Z^n) + I (U_1^n;Y_1^n|U_2^n,Z^n)\nonumber\\
&\le& I(U_1^n;U_2^n,Z^n)+ I(X_1^n;Y_1^n|U_2^n,Z^n)\label{ap2}.
\end{eqnarray}

But,
\begin{eqnarray}
I(X_1^n;Y_1^n|U_2^n,Z^n)&=& H(Y_1^n|U_2^n,Z^n)- H (Y_1^n|X_1^n)\le H(Y_1^n)- H (Y_1^n|X_1^n)\nonumber\\
&\le& \sum_{i=1}^n H(Y_{1i})-\sum_{i=1}^n H(Y_{1i}|Y_1^{i-1},X_{1i})= \sum_{i=1}^n H(Y_{1i})-\sum_{i=1}^nH(Y_{1i}|X_{1i})\nonumber\\
&=&\sum_{i=1}^n I(X_{1i};Y_{1i})\label{ap3}.
\end{eqnarray}
The inequality in the second line is due to the fact that conditioning reduces entropy and the equality in the fifth line is due to the memoryless property of the channel.

From \eqref{ap1}, \eqref{ap2} and \eqref{ap3}
\begin{eqnarray*}
H(U_1) &\le& \frac{1}{n} \sum_{i=1}^n I (U_{1i};U_{2i},Z_i) + \frac{1}{n}
\sum_{i=1}^n I(X_{1i};Y_{1i})+ \lambda_n.
\end{eqnarray*}
We can introduce time sharing random variable as done in
~\cite{Cover04elements} and show that  $H(U_1) \le I(U_1;U_2,Z)+ I(X_1;Y_1)$. This simplifies to
$H(U_1|U_2,Z) \le I(X_1;Y_1)$.

 By the symmetry of the problem we get $H(U_2|U_1,Z) \le I(X_2;Y_2)$.

We also have
\begin{eqnarray*}
n H(U_1,U_2)&=&H(U_1^n,U_2^n)\\
&=& H(U_1^n,U_2^n|Y_1^n,Y_2^n,Z^n)+ I (U_1^n,U_2^n;Y_1^n,Y_2^n,Z^n)\\
&\le& I(U_1^n,U_2^n; Y_1^n,Y_2^n,Z^n)+n \lambda_n.
\end{eqnarray*}

But
\begin{eqnarray*}
I(U_1^n,U_2^n; Y_1^n,Y_2^n,Z^n)&=& I(U_1^n,U_2^n;,Z^n) + I (U_1^n,U_2^n;Y_1^n,Y_2^n|Z^n)\\
&\le& I(U_1^n,U_2^n;Z^n)+ I(X_1^n,X_2^n;Y_1^n,Y_2^n|Z^n).
\end{eqnarray*}
Also,
\begin{eqnarray*}
I(X_1^n,X_2^n;Y_1^n,Y_2^n|Z^n)&=& H(Y_1^n,Y_2^n|Z^n)- H (Y_1^n,Y_2^n|X_1^n,X_2^n,Z^n)\\
&\le& H(Y_1^n,Y_2^n)- H (Y_1^n|X_1^n)-H(Y_2^n|X_2^n)\\
&\le& H(Y_1^n) + H(Y_2^n)- H (Y_1^n|X_1^n)-H(Y_2^n|X_2^n).
\end{eqnarray*}
Then, following the  steps used above, we obtain $H(U_1,U_2|Z) \le I(X_1;Y_1)+ I (X_2;Y_2)$.~~~~~~~~~~~~~~~~~~~~~~~~~~~$\blacksquare$

\section{Proofs of Lemmas in Section 4}\label{l1}

{{\bf{Proof of Lemma 1}}}:  Let  
   $\Delta\buildrel\Delta\over = I ( X_1;Y | X_2, U_2) -  I ( X_1;Y | X_2 )$. Then denoting  differential entropy by  $h$,
\begin{eqnarray}
\Delta= h(Y | X_2, U_2) - h(Y | X_1, X_2, U_2)  - [h(Y | X_2) - h (Y | X_1, X_2)].\nonumber
\end{eqnarray}
Since the channel is memoryless, $ h(Y | X_1, X_2, U_2) = h(Y | X_1, X_2)$.   Thus,  $\Delta  \leq 0$.

~~~~~~~~~~~~~~~~~~~~~~~~~~~~~~~~~~~~~~~~~~~~~~~~~~~~~~~~~~~~~~~~~~~~~~~~~~~~~~~~~~~~~~~~~~~~~~~~~~~~~~~~~~~~~~~~~~~~~~~~~~~~~~~$\blacksquare$

{\bf{Proof of Lemma 2}}: Since
\begin{equation*}
 I (X_1, X_2;Y) = h(Y) - h(Y | X_1, X_2)= h (X_1+ X_2+ N) - h(N),
\end{equation*}
it is maximized when $h (X_1+ X_2+ N)$ is maximized. This entropy is maximized when $X_1+ X_2$ is Gaussian with the largest possible variance $=P_1+P_2$. If $(X_1, X_2)$ is jointly Gaussian then so is $X_1+ X_2$.

Next consider $I (X_1;Y | X_2 )$. This equals 
\begin{equation*}
h (Y | X_2) - h (N)=  h (X_1+ X_2+ N | X_2) - h (N)= h (X_1+ N | X_2) - h(N)
\end{equation*}
which is maximized when $p (x_1| x_2)$ is Gaussian and this happens when $X_1, X_2$ are jointly Gaussian.	
			   
A similar result holds for $I (X_2; Y | X_1)$.
~~~~~~~~~~~~~~~~~~~~~~~~~~~~~~~~~~~~~~~~~~~~~~~~~~~~~~~~~~~~~~~~~~~~~~~~~~~~~~~~~~~~$\blacksquare$

{{\bf{Proof of Lemma 3}}}: Since $X_1\leftrightarrow U_1\leftrightarrow U_2\leftrightarrow X_2$  is a Markov chain, by data processing inequality
$I (X_1; X_2) \leq I (U_1; U_2)$. Taking $X_1, X_2$ to be jointly Gaussian with zero mean, unit variance and correlation ${\rho},~I(X_1,X_2)=0.5log_2(\frac{1}{1-{\rho}^2})$.  This implies $ {\rho}^2 \leq 1- 2^{-2I(U_1,U_2)}$.
~~~~~~~~~~~~~~~~~~~~~~~~~~~~~~~~~~~~~~~~~~~~~~~~$\blacksquare$

{{\bf{Proof of Lemma 4}}}: By Stone-Weierstrass theorem (\cite{protter}, \cite{roy}) the class of functions $(x_1,x_2) \mapsto e^{\frac{-1}{2c_{1}}(x_1-a_{1})^2}$$ e^{\frac{-1}{2c_{2}}(x_2-a_{2})^2}$ can be shown to be dense in $C_0$ under uniform convergence where $C_0$ is the set of all continuous functions on $\Re^2$ such that $\lim_{\|X\| \to \infty}|f(x)|=0$  . Since the jointly Gaussian density $(x_1,x_2) \mapsto e^{\frac{-1}{2\sigma^2}(\frac{x_1^2+x_2^2-2\rho x_1 x_2}{1-\rho^2})}$  is in $C_0$, it can be approximated arbitrarily closely uniformly by the functions \eqref{co2}.~~~~~~~~~~~~~~~~~~~~~~~~~~~~~~~~~~~~~~~~~~~~~~~~~~~~~~~~~~~~~~~~~~~~~~~~~~~~~~~~~~~~~~~~~~~~~~~~~~~~~~~~~~~~~~~~~~~~~~$\blacksquare$

{{\bf{Proof of Lemma 5}}}:  Since
\begin{equation*} 
I(X_{m1},X_{m2};Y_m)= h(Y_m)-h(Y_m|X_{m1},X_{m2})= h(Y_m)-h(N),
\end{equation*}
it is sufficient to show that $h(Y_m) \rightarrow h(Y)$.
From $(X_{m1},X_{m2}) {\buildrel d \over \longrightarrow} (X_1,X_2)$  and independence of $(X_{m1},X_{m2})$   from $N$, we get 
 $ Y_m= X_{m1}+ X_{m2} + N {\buildrel d \over \longrightarrow} X_1+X_2+N = Y$. Then $f_m \rightarrow f$  uniformly implies that $f_m(Y_m){\buildrel d \over \longrightarrow} f(Y)$. Since $f_m(Y_m) \geq 0,~ f(Y) \geq 0~ a.s $  and $log$ being continuous except at $0$, we  obtain $ logf_m(Y_m){\buildrel d \over \longrightarrow} logf(Y)$. Then uniform integrability provides $I(X_{m1},X_{m2};Y_m) \rightarrow I(X_1,X_2;Y)$.~~~~~~~~~$\blacksquare$

\bibliographystyle{abbrv}
\bibliography{mybibfile_chap}

\end{document}